\documentclass{emulateapj}   
\usepackage{apjfonts}

\submitted{ApJ, accepted}

\def\ltsima{$\; \buildrel < \over \sim \;$}
\def\simlt{\lower.5ex\hbox{\ltsima}} 
\def\gtsima{$\; \buildrel > \over \sim \;$}
\def\simgt{\lower.5ex\hbox{\gtsima}} 

\def\deg{\hbox{$^\circ$}}
\def\gunit{$\times$ 10$^{-8}$ ph cm$^{-2}$ s$^{-1}$}

\shorttitle{Fermi-LAT $\gamma$-ray Detection of M87}
\shortauthors{Fermi-LAT Collaboration}

\begin{document}

\title{Fermi Large Area Telescope Gamma-Ray Detection of the Radio Galaxy M87}

\author{
A.~A.~Abdo\altaffilmark{2,3}, 
M.~Ackermann\altaffilmark{4}, 
M.~Ajello\altaffilmark{4}, 
W.~B.~Atwood\altaffilmark{5}, 
M.~Axelsson\altaffilmark{6,7}, 
L.~Baldini\altaffilmark{8}, 
J.~Ballet\altaffilmark{9}, 
G.~Barbiellini\altaffilmark{10,11}, 
D.~Bastieri\altaffilmark{12,13}, 
K.~Bechtol\altaffilmark{4}
% ...
\L.~Stawarz\altaffilmark{56,4}
}
% ...
\altaffiltext{2}{Space Science Division, Naval Research Laboratory, Washington, DC 20375, USA}
\altaffiltext{3}{National Research Council Research Associate, National Academy of Sciences, Washington, DC 20001, USA}
\altaffiltext{4}{W. W. Hansen Experimental Physics Laboratory, Kavli Institute for Particle Astrophysics and Cosmology, Department of Physics and SLAC National Accelerator Laboratory, Stanford University, Stanford, CA 94305, USA}
\altaffiltext{5}{Santa Cruz Institute for Particle Physics, Department of Physics and Department of Astronomy and Astrophysics, University of California at Santa Cruz, Santa Cruz, CA 95064, USA}
\altaffiltext{6}{Department of Astronomy, Stockholm University, SE-106 91 Stockholm, Sweden}
\altaffiltext{7}{The Oskar Klein Centre for Cosmoparticle Physics, AlbaNova, SE-106 91 Stockholm, Sweden}
\altaffiltext{8}{Istituto Nazionale di Fisica Nucleare, Sezione di Pisa, I-56127 Pisa, Italy}
\altaffiltext{9}{Laboratoire AIM, CEA-IRFU/CNRS/Universit\'e Paris Diderot, Service d'Astrophysique, CEA Saclay, 91191 Gif sur Yvette, France}
\altaffiltext{10}{Istituto Nazionale di Fisica Nucleare, Sezione di Trieste, I-34127 Trieste, Italy}
\altaffiltext{11}{Dipartimento di Fisica, Universit\`a di Trieste, I-34127 Trieste, Italy}
\altaffiltext{12}{Istituto Nazionale di Fisica Nucleare, Sezione di Padova, I-35131 Padova, Italy}
\altaffiltext{13}{Dipartimento di Fisica ``G. Galilei", Universit\`a di Padova, I-35131 Padova, Italy}
% ...
\altaffiltext{56}{Astronomical Observatory, Jagiellonian University, 30-244 Krak\'ow, Poland}

\begin{abstract}

We report the $Fermi$-LAT discovery of high-energy (MeV/GeV) 
$\gamma$-ray emission positionally consistent with the center of the 
radio galaxy M87, at a source significance of over $10\sigma$ in 
ten-months of all-sky survey data. Following the detections of Cen~A and 
Per~A, this makes M87 the third radio galaxy seen with the LAT. The 
faint point-like $\gamma$-ray source has a $>$100 MeV flux of 2.45 ($\pm 
0.63$) \gunit\ (photon index = $2.26 \pm 0.13$) with no significant 
variability detected within the LAT observation. This flux is comparable 
with the previous EGRET upper limit ($< 2.18$ \gunit, 2$\sigma$), thus 
there is no evidence for a significant MeV/GeV flare on decade 
timescales. Contemporaneous $Chandra$ and VLBA data indicate low 
activity in the unresolved X-ray and radio core relative to previous 
observations, suggesting M87 is in a quiescent overall level over the 
first year of $Fermi$-LAT observations. The LAT $\gamma$-ray spectrum is 
modeled as synchrotron self-Compton (SSC) emission from the electron 
population producing the radio-to-X-ray emission in the core. The 
resultant SSC spectrum extrapolates smoothly from the LAT band to the 
historical-minimum TeV emission. Alternative models for the core and 
possible contributions from the kiloparsec-scale jet in M87 are 
considered, and can not be excluded.

\end{abstract}


\section{Introduction}\label{section-intro}

As one of the nearest radio galaxies to us \citep[$D=16$ Mpc is 
adopted;][]{ton91}, M87 is amongst the best-studied of its source class. 
It is perhaps best known for its exceptionally bright arcsecond-scale 
jet \citep{cur18}, well-imaged at radio through X-ray frequencies at 
increasingly improved sensitivity and resolution over the decades 
\citep[e.g.,][]{bir91,spa96,mar02,per05}. Near its central $\sim (3-6) 
\times 10^{9}$ solar mass supermassive black hole \citep{mac97,geb09}, 
the jet base has been imaged down to $\sim$0.01 pc resolution 
\citep[$\sim 15-30\times$ the Schwarzschild 
radius,][]{jun99,ly07,tev09}.

At the highest energies, M87 is regularly detected by HESS, MAGIC, and 
VERITAS with variable TeV emission on timescales of years and flaring in 
a few days \citep{aha06,alb08,acc08,tev09}. The sensitivity of these 
Cherenkov telescopes have also enabled the detection of another 
well-known nearby radio galaxy Cen~A \citep{aha09}. Without comparable 
imaging resolution to the lower energy studies however, variability and 
spectral modeling are necessary to infer the production site of the TeV 
$\gamma$-rays and to deduce the source physical parameters.

At high energy $\gamma$-rays ($\sim$20 MeV -- 100 GeV), we are similarly 
poised for new radio galaxy discoveries with the Large Area Telescope (LAT) 
aboard the recently launched {\it Fermi Gamma-ray Space Telescope} 
\citep{atw09}. Indeed, we report here the detection of a faint, point-like 
$\gamma$-ray source positionally coincident with M87 using the $Fermi$-LAT. 
After the confirmation of the EGRET discovery of Cen~A \citep[][and in 
preparation]{sre99,lbas}, and the recent detection of Per~A/NGC~1275 
\citep{pera}, this is the third radio galaxy successfully detected by $Fermi$. 
Unlike the known variable TeV source, there is so far no evidence for 
variability of the MeV/GeV emission in M87. An origin of the LAT emission from 
the unresolved parsec scale jet (hereafter, denoted as the `nucleus' or 
`core') observed contemporaneously with $Chandra$ and the VLBA\footnote{The 
National Radio Astronomy Observatory is a facility of the National Science 
Foundation operated under cooperative agreement by Associated Universities, 
Inc.} is discussed. Potential contributions from the larger-scale ($\simgt 
0.1-1$ kpc) jet to the unresolved $\gamma$-ray source are also briefly 
considered. Section 2 contains the details of the LAT observations, including 
a description of the $Chandra$ and VLBA data utilized, with the discussion 
of these results in section 3.

\section{Observations}\label{section-lat}

The $Fermi$-LAT is a pair creation telescope which covers the energy range from 
$\sim$20 MeV to $>$300 GeV \citep{atw09}. It operates primarily in an `all-sky 
survey' mode, scanning the entire sky approximately every three hours.
The initial LAT detection of M87 resulted from nominal processing of 
6-months of all-sky survey data, as was applied to the initial 3-month 
dataset described in \citet{bsl}, with a test statistic \citep{mat96}, 
$TS \sim 60$. Including here an additional 4-months of data, the $TS$ 
increased to 108.5, which is equivalent to a source significance 
$\sim\sqrt{TS}=10.4\sigma$. The resultant 10-month dataset (Aug.\ 4, 
2008 - May 31, 2009) corresponds to mission elapsed times (MET) 
239557418 to 265420800. Our analysis followed standard selections of 
"Diffuse" class events \citep{atw09} with energies $E>$200 MeV, a zenith 
angle cut of $<$105\deg, and a rocking angle cut of 43\deg\ applied in 
order to avoid Earth albedo $\gamma$-rays. {\it Fermi} Science 
tools\footnote{http://fermi.gsfc.nasa.gov/ssc/data/analysis/documentation/Cicerone/} 
version v9r10 and instrumental response functions (IRFs) version {\tt 
P6$\_$V3$\_$DIFFUSE} were used for the analysis.

\begin{figure}
\epsscale{1.1}
\plotone{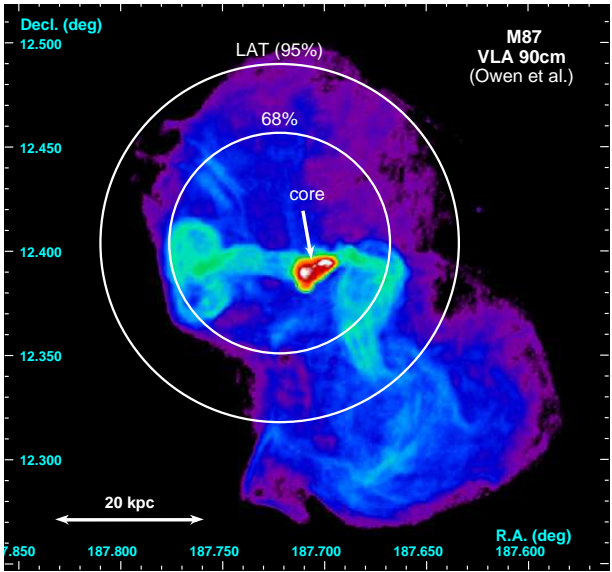}
 \caption{VLA $\lambda$=90cm radio image from \citet{owe00} with the LAT
$\gamma$-ray localization error circles indicated: $r_{95\%}= 5.2'$ and
$r_{68\%}=3.2'$ (statistical only; see \S~\ref{section-lat}). The M87
core is the faint feature near the center of the few kpc-scale
double-lobed radio structure (in white). At the adopted distance $D=16$
Mpc, 1$'$ = 4.7 kpc. 
}
\label{figure-vla}
\end{figure}

A localization analysis with {\tt GTFINDSRC} resulted in a best-fit 
position, RA = 187\deg.722, Dec.\ = 12\deg.404 (J2000.0 equinox), with a 
95$\%$ confidence error radius, $r_{95\%}= 0\deg.086 = 5.2'$ 
(statistical only; $r_{68\%}=3.2'$). To account for possible 
contamination from nearby sources, the model included all point sources 
detected at $>5\sigma$ in an internal LAT 9-month source list within a 
region of interest (ROI) of $r$=15\deg\ centered on the $\gamma$-ray 
position. Galactic diffuse emission was modeled using GALPROP 
\citep{str04}, updated to include recent gas maps and a more accurate 
decomposition into Galactocentric rings (galdef ID {\tt 
54$\_$59varh7S}). An additional isotropic diffuse component modeled as a 
power-law was included. Figure~\ref{figure-vla} shows the resultant 
$\gamma$-ray source localization on a VLA radio image from 
\citet{owe00}. The $\gamma$-ray source is positionally coincident with 
the known radio position of the M87 core \citep[RA = 187\deg.706, Dec.\ 
= 12\deg.391;][]{fey04}, with an offset (0\deg.020 = 1.2$'$) that is a 
small fraction of the localization circle. Currently, the best estimate 
of the systematic uncertainty in $r_{95\%}$ is 2.4$'$ \citep{bsl}, which 
should be added in quadrature to the determined statistical one.

\begin{figure}
\epsscale{1.2}
\plotone{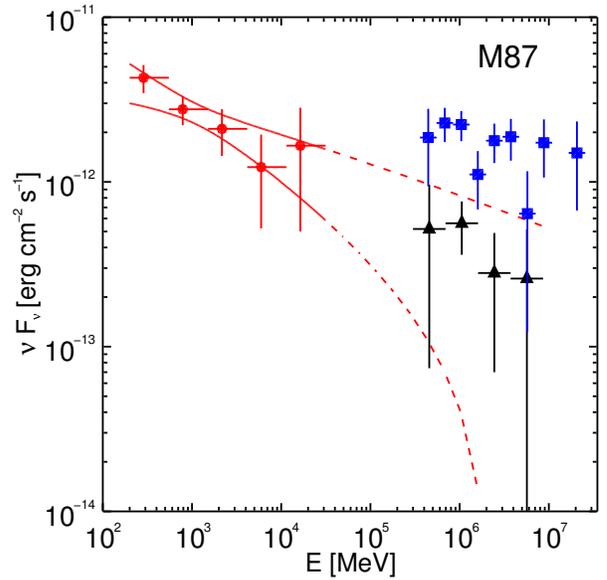}
 \caption{The observed LAT spectrum (red circles) with representative
TeV measurements of M87 in a low state from the 2004 observing season
(black triangles) and during a high state in 2005 (blue squares), both
by HESS \citep{aha06}. The lines indicate 1$\sigma$ bounds on the power
law fit to the LAT data as well as its extrapolation into TeV energies.
}
\label{figure-sedgamma}
\end{figure}

Spectral analysis was performed utilizing an unbinned likelihood fit of 
the $>$200 MeV data with a power-law ($dN/dE \propto E^{-\Gamma}$) 
implemented in the {\tt GTLIKE} tool. This resulted in $F$($>$100 MeV) = 
2.45 ($\pm 0.63$) \gunit\ with a photon index, $\Gamma = 2.26 \pm 0.13$; 
errors are statistical only. The flux was extrapolated down to 100 MeV 
to facilitate comparison with the previous EGRET non-detection of $<$ 
2.18 \gunit\ (2$\sigma$) from observations spanning the 1990's 
\citep{rei03}. Thus, there is no apparent changes in the flux (i.e., a 
rise) in the decade since the EGRET observations. Systematic errors of 
($+0.17$/$-0.15$) \gunit\ on the flux and $+0.04$/$-0.11$ on the index 
were derived by bracketing the energy-dependent ROI of the IRFs to 
values of 10$\%$, 5$\%$, and 20$\%$ above and below their nominal values 
at log($E$[MeV]) = 2, 2.75, and 4, respectively. The spectrum extends to 
just over 30 GeV where the highest energy photon is detected within the 
95$\%$ containment.  The LAT spectral data points presented in 
Figure~\ref{figure-sedgamma} were generated by performing a subsequent 
likelihood analysis in five equal logarithmically spaced energy bins 
from $0.2-31.5$ GeV. The 1$\sigma$ bounds on the spectrum, obtained from 
the full $>$200 MeV unbinned likelihood fit, were extended to higher 
energies for comparison with previous TeV measurements (see section~3).

\begin{figure}
\epsscale{1.2}
\plotone{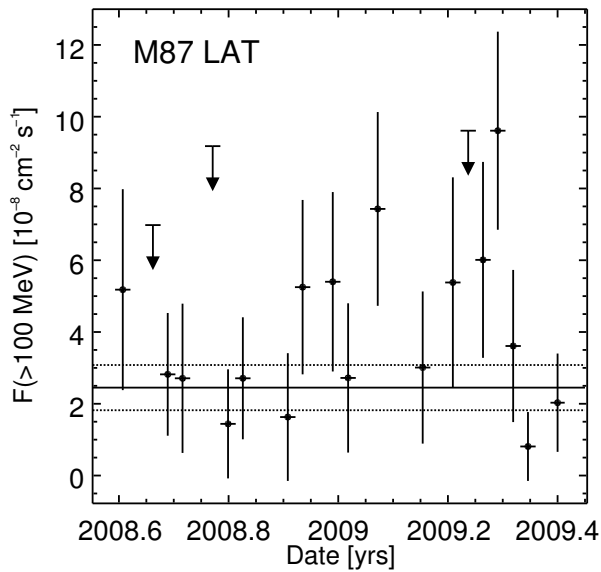}
 \caption{Lightcurve in 10-day bins obtained with the fitted photon
index ($\Gamma$=2.26) fixed. The average flux is indicated with the
solid horizontal line and the dotted lines are $\pm 1 \sigma$ about the
average. Data points with $TS<1$ (i.e., 1$\sigma$) are shown as upper
limits.
}
\label{figure-lightcurve}
\end{figure}

Lightcurves were produced in 10-day (Figure~\ref{figure-lightcurve}) and 
28-day (not shown) bins over the 10-month LAT dataset. Considering the 
limited statistics, it was necessary to fix the photon index to the 
(average) fitted value in order to usefully gauge variability in the 
flux. Considering only statistical errors of all the binned data points 
with $TS \geq1$ ($1\sigma$), a $\chi^{2}$ test against the weighted mean 
fluxes of the 10-day and 28-day lightcurves resulted in probabilities, 
$P(\chi^{2},\nu) = 22\%$ and 70$\%$, respectively, indicating plausible 
fits to the tested hypothesis.  We conclude that there is no evidence 
for variability over the period of observations.

A radial profile of the $\gamma$-ray source counts (not shown) was 
extracted for the total energy range ($>$200 MeV). The profile is 
consistent with that of a point source simulated at energies $0.2-200$ 
GeV using the fitted spectral parameters above with a reduced 
$\chi^{2}=1.04$ for 20 degrees of freedom. The total $\sim$0\deg.2 
extent of the 10's kpc-scale radio lobes of M87 
\citep[Figure~\ref{figure-vla};][]{owe00} is comparable to the LAT 
angular resolution, $\theta_{\rm 68} \simeq 0\deg.8~ E_{\rm GeV}^{-0.8}$ 
\citep{atw09}. Therefore, from the presently available data, we can not 
disentangle (or exclude) a possible contribution of the extended radio 
features to the total $\gamma$-ray flux.

To gauge the X-ray activity of M87 over the duration of the LAT 
observations, we analyzed five new 5 ksec $Chandra$ ACIS-S images 
obtained in $\sim$6 week intervals between Nov.\ 2008 and May 2009 (PI: 
D.~E.\ Harris). The X-ray core fluxes (0.5--7 keV) in these monitoring 
observations, $(1.2-1.6) \times 10^{-12}$ erg cm$^{-2}$ s$^{-1}$ 
\citep[$\sim 0.4-0.6$ keV s$^{-1}$ in the units of Fig.~9 of][]{har09}, 
are at the low end of the observed range over the last $\sim$7 years. 
Additionally, the fractional variability is small ($\sigma$/$<$flux$> 
\sim 0.1$), indicating low X-ray activity in the core over the LAT 
observing period.

At milli-arcsecond (mas) resolution in the radio band, M87 has been 
monitored with the VLBA at 15 GHz since 1995 as part of the 2cm survey 
\citep{kel04} and MOJAVE \citep{lis09} programs\footnote{See: 
http://www.physics.purdue.edu/MOJAVE/}. These data were re-imaged 
uniformly at 0.6 mas $\times$ 1.3 mas (position angle=--11\deg) 
resolution to match the additional map presented in \citet{kov07} 
resulting in 23 total measurements of the unresolved core flux up to the 
latest observation on Jan.\ 7, 2009 (one of the 
$Chandra$ exposures described above was obtained on the same day). 
This observation is the only one overlapping with the LAT dataset 
and the peak flux of 1.05 Jy/beam is consistent with the average over 
all the measurements ($1.11 \pm 0.16$ Jy/beam). An indication of the 
sensitivity of these data to detecting flaring core emission is that the 
high flux state observed in the detailed 43 GHz VLBA monitoring at the 
time of TeV flaring in 2008 \citep[][see 
\S~\ref{section-discussion}]{tev09} is visible in the 15 GHz data as a 
single high point on May 1, 2008 (1.45 Jy/beam)\footnote{Conversely, 
during the previous TeV flaring period spanning March - May 2005 
\citep{aha06}, no comparable flare was visible in the VLBA 15 GHz core: 
1.02 Jy/beam peak on Apr.\ 21st and 0.98 Jy/beam on Nov.\ 7th. Instead, 
the flaring X-ray/optical/radio knot HST-1 (60 pc from the core, 
projected) was observed to peak in early 2005 \citep{har09}, suggesting 
an association with the variable TeV source 
\citep{che07}.\label{footnote-hst1}}. This only suggests a period of low 
activity in the radio core (as reflected in the X-ray data), as the 
single radio flux may not be representative of the entire 10-month LAT 
viewing period.

\section{Discussion}\label{section-discussion}

In blazars, it is commonly believed that $\gamma$-rays are produced in 
compact emission regions moving with relativistic bulk velocities in or 
near the parsec scale core in order to explain the observed rapid 
variability and to avoid catastrophic pair-production 
\citep[e.g.,][]{don95}. Consequently, it is natural to extend this 
supposition to radio galaxies \citep{chi01} which are believed to have 
jets oriented at systematically larger angles to our line of sight, thus 
constituting the parent population of blazars. Indeed, in the case of 
M87, a significant months timescale rise in the flux of the sub-pc scale 
radio core was discovered with the VLBA (at 43 GHz) during a period in 
early 2008 when few day timescale TeV flaring was detected 
\citep{tev09}. During this period of increased activity, a $Chandra$ 
measurement of the sub-arcsecond scale X-ray nucleus also indicated a 
relatively higher flux than seen in past observations \citep{har09}, 
thus signaling a common origin for the flaring emissions in the M87 
nucleus. Therefore, during periods of lower $\gamma$-ray activity, the 
radio/X-ray core can also be considered a dominant source of the 
unresolved higher-energy emission, and we discuss this in the context of 
the LAT MeV/GeV detection.

In Figure~\ref{figure-sedgamma}, the LAT spectrum of M87 is plotted 
along with representative integrated TeV spectra from HESS 
\citep{aha06}. The TeV measurements cover periods when M87 was in its 
historical-minimum (in 2004), and during a high state \citep[in 2005 -- 
cf., Fig.~3 in][]{acc08}. Although the formal difference in the fitted 
photon indices of the TeV data at high and low states is not 
statistically significant ($\Gamma = 2.22 \pm 0.15$ and $2.62 \pm 0.35$, 
respectively), the LAT MeV/GeV spectrum ($\Gamma =2.3$) connects 
smoothly with the low-state TeV spectrum. Taken together with the X-ray 
and radio measurements obtained during the LAT observation 
(\S~\ref{section-lat}), we view this as an indication that M87 is in an 
overall low $\gamma$-ray activity state during the considered period. In 
fact, no significant TeV flaring was detected in a preliminary analysis 
of 18 hrs of contemporaneous VERITAS observations from Jan.\ - Apr.\ 
2009 \citep{hui09}.

M87 is the faintest $\gamma$-ray radio galaxy detected so far by the LAT 
with a $>$100 MeV flux ($\sim 2.5$\gunit) about an order of magnitude 
lower than in Cen~A \citep{lbas} and Per~A \citep{pera}; the 
corresponding $>100$ MeV luminosity, $4.9 \times 10^{41}$ erg s$^{-1}$, 
is 4$\times$ greater than that of Cen~A, but $>$200$\times$ smaller than 
in Per A. There is no evidence of intra-year or decade-timescale MeV/GeV 
variability in M87 (\S~\ref{section-lat}), in contrast to the $\simgt 
7\times$ and $\sim 1.6\times$ larger observed LAT fluxes than the 
previous EGRET ones in the cases of Per~A \citep{pera} and Cen~A 
\citep{lbas}, respectively. The $\gamma$-ray photon index of M87 in the 
LAT band is similar to that of Per~A ($\Gamma = 2.3$ and 2.2, 
respectively), while being smaller than observed in Cen~A 
\citep[$\Gamma=2.9$;][]{lbas}. These sources are low-power (FRI) radio 
galaxies, and have broad low-energy synchrotron and high-energy inverse 
Compton (IC) components in their spectral energy distributions (SEDs) 
peaking roughly in the infrared and $\gamma$-ray bands, respectively. 
Low-energy peaked BL Lac objects have similar shaped SEDs, with 
approximately equal apparent luminosities \citep[e.g.,][]{kub98}. As FRI 
radio galaxies are believed to constitute the parent population of BL 
Lacs in unified schemes \citep{urr95}, the overall similarity of their 
SEDs is not surprising.

\begin{figure}
\epsscale{1.2}
\plotone{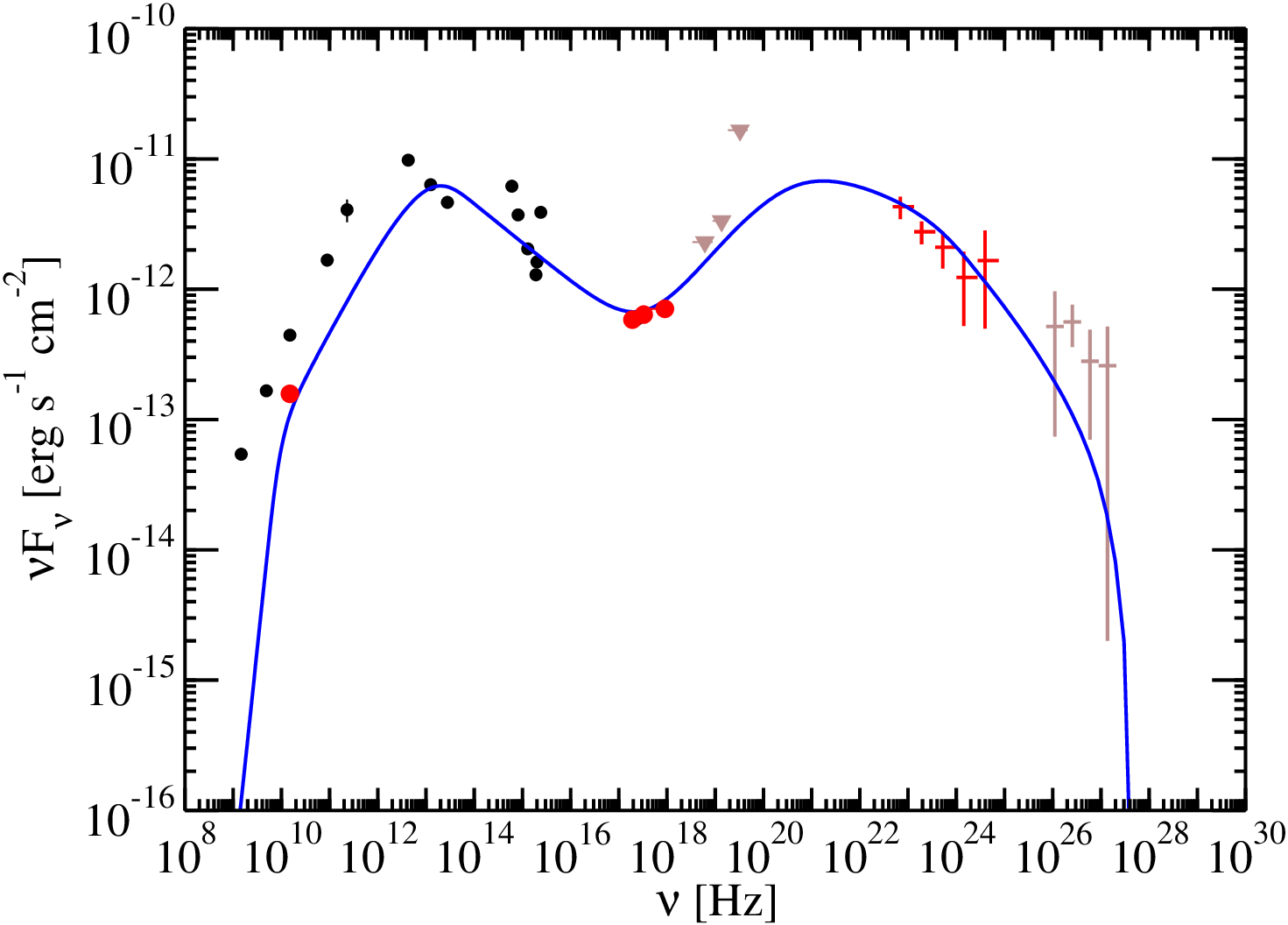}
 \caption{SED of M87 with the LAT spectrum and the Jan.\ 7, 2009 MOJAVE
VLBA 15 GHz and $Chandra$ X-ray measurements of the core indicated in
red. The non-simultaneous 2004 TeV spectrum described in
Figure~\ref{figure-sedgamma} and $Swift$/BAT hard X-ray limits
(\S~\ref{section-discussion}) of the integrated emission are shown in
light brown. Historical measurements of the core from VLA 1.5, 5, 15 GHz
\citep{bir91}, IRAM 89 GHz \citep{des96}, SMA 230 GHz \citep{tan08},
$Spitzer$ 70, 24 $\mu$m \citep{shi07}, Gemini 10.8 $\mu$m \citep{per01},
$HST$ optical/UV \citep{spa96}, and $Chandra$ 1 keV from \citet[][hidden
behind the new measurements]{mar02} are plotted as black circles. The
VLBA 15 GHz flux is systematically lower than the historical
arcsec-resolution radio to infrared measurements due to the presence of
intermediate scale emission \citep[see e.g.,][]{kov07}. The blue line
shows the one-zone SSC model fit for the core described in
\S~\ref{section-discussion}.
}
\label{figure-sed}
\end{figure}

We construct a SED for M87 (Figure~\ref{figure-sed}) using the the LAT 
$\gamma$-ray spectrum and the overlapping Jan.\ 7, 2009 $Chandra$ and 
VLBA measurements of the core. Also plotted are historical radio to 
X-ray fluxes of the core \citep[see][]{spa96,tan08} measured at the 
highest resolutions at the respective frequencies. The core is known to 
be variable, with factors of $\sim2$ changes on months timescales common 
in the optical and X-ray bands \citep{per03,har09}. To help constrain 
the overall SED at frequencies between the X-ray and LAT measurements, 
we determined integrated 3$\sigma$ upper limits in three hard X-ray 
bands \citep[following,][]{aje08} based on the $Swift$/BAT dataset in 
\citet{aje09}, including about another additional year of exposure 
(i.e., $\sim$4 years total from Mar.\ 2005 - Jan.\ 2009).

The broad-band SED is fit with a homogeneous one-zone synchrotron 
self-Compton (SSC) jet model \citep{fin08} assuming an angle to the line 
of sight, $\theta$=10\deg, and bulk Lorentz factor, $\Gamma_{\rm b} = 
2.3$ (Doppler factor, $\delta = 3.9$), consistent with observations of 
apparent motions of $\simgt 0.4c$ ($\Gamma_{\rm b} > 1.1$) in the 
parsec-scale radio jet \citep{ly07}. A broken power-law electron energy 
distribution $N(\gamma)\propto\gamma^{-p}$ is assumed, and the indices, 
$p_{\rm 1}=1.6$ for $\gamma$ = [1, $4 \times 10^{3}$] and $p_{\rm 
2}=3.6$ for $\gamma$ = [$4 \times 10^{3}$, $10^{7}$] are best guesses 
based on the available core measurements. The normalization at low 
energies is constrained by the single contemporaneous VLBA 15 GHz flux 
which is measured with $\sim 10^{2}-10^{3}\times$ better resolution than 
the adjacent points. The source radius, $r= 1.4 \times 10^{16}$ cm = 4.5 
mpc is chosen to be consistent with the best VLBA 43 GHz map resolution 
\citep[$r<$7.8 mpc = 0.1 mas,][]{jun99,ly07} and is of order the size 
implied by the few day timescale TeV variability \citep{tev09}. For the 
source size adopted, internal $\gamma-\gamma$ absorption is avoided so 
that the LAT spectrum extends relatively smoothly into the TeV band, 
consistent with the historical-minimum flux detected by HESS 
\citep{aha06} and the preliminary upper limit of $<1.9\%$ Crab from 
VERITAS observations \citep{hui09} contemporaneous with the LAT ones.

In the SSC model, the magnetic field is $B = 55$ mG 
and assuming the proton energy density is 10$\times$ greater than the 
electron energy density, the total jet power is $P_{\rm j} \sim 7.0 
\times 10^{43}$ erg s$^{-1}$. The jet power is particle dominated, with 
only a small contribution from the magnetic field component ($P_{\rm B} 
\sim 2 \times 10^{40}$ erg s$^{-1}$). In comparison, the total kinetic 
power in the jet is $\sim$few $\times 10^{44}$ erg s$^{-1}$ as 
determined from the energetics of the kpc-scale jet and lobes 
\citep{bic96}, and is consistent with the jet power available from 
accretion, $P_{\rm j} \simlt 10^{45}$ erg s$^{-1}$ \citep{rey96,dim03}. 
These power estimates are similar to those derived for BL Lacs from 
similarly modeling their broad-band SEDs \citep[e.g.,][]{cel08}.

As applied to M87, such single-zone SSC emission models also reproduce 
well the broad-band SEDs up to MeV/GeV energies in the radio galaxies 
Per~A \citep{pera} and Cen~A \citep{chi01}. In this context, the 
observed MeV/GeV $\gamma$-ray fluxes of blazars appear to be correlated 
with their compact radio cores \citep{lbas,kov09}, suggesting a common 
origin in the Doppler boosted emission in the sub-parsec scale jets. The 
fact that the 3 radio galaxies detected by the LAT so far have amongst 
the brightest ($\simgt$1 Jy) unresolved radio cores, in line with these 
expectations \citep{ghi05}, lend evidence for a common connection 
between the $\gamma$-ray and radio emitting zones in such jets.

It should be emphasized that these observations are not simultaneous and 
particularly, the TeV emission is known to be variable on year 
timescales, so other emission components may contribute to the variable 
emission. Therefore, although not strictly required, more sophisticated 
models over the single-zone one presented can reproduce or contribute to 
the observed emission. In particular, the beaming requirements in the 
one-zone SSC modeling of the three known $\gamma$-ray radio galaxies are 
systematically lower than required in BL Lacs, suggesting velocity 
profiles in the flow \citep{chi01}. Such models \citep{geo05,tav08} have 
in fact been used to fit the SED of M87 in addition to models based on 
additional spatial structure \citep[e.g.,][]{len07}. Protons, being 
inevitably accelerated if they co-exist with electrons in the emission 
regions, probably dominate energetically and dynamically the jets of 
powerful AGN \citep[e.g.,][]{cel08}. Applying the synchrotron-proton 
blazar model \citep{muc03,rei04} to the quiescent M87 data set yields 
reasonable agreement model fits that support a highly magnetized compact 
emission region with approximate equipartition between fields and 
particles and a total jet power comparable with the above estimates, 
where protons are accelerated up to $\sim 10^{9}$ GeV.

Outside of the pc-scale core, the well-known arcsecond-scale jet 
\citep[e.g.,][]{bir91,mar02,per05} is also a possible source of IC 
emission. As both the LAT and TeV telescopes are unable to spatially 
resolve emission on such small scales, the expected spectral and 
temporal properties of the predicted emission must be examined. On the 
observed scales, the dominant seed photon source is the host galaxy 
starlight, and such an IC/starlight model applied to one of the 
brightest resolved knots in the jet -- knot A, $\sim$1 kpc projected 
distance from the core -- results in a spectrum peaking at TeV energies 
\citep{sta05}, thus producing a harder MeV/GeV spectrum than observed by 
the LAT. Even closer to the core ($\sim$60 pc, projected), the 
superluminal knot HST-1 \citep[$>4c-6c$;][]{bir99,che07} is a more 
complex case. This knot is more compact than knot A, and its IC emission 
is expected to be further enhanced by the increased energy densities of 
the surrounding circumnuclear and galactic photon fields, as well of the 
comoving synchrotron radiation \citep{sta06}. The radio/optical/X-ray 
fluxes of HST-1 have been declining since its giant flare peaked in 2005 
\citep{har09}, with current X-ray fluxes comparable to its pre-flare 
levels in 2002. Considering the variable and compact nature of the 
source (with observed months doubling timescales implying $r\simlt 22 
\delta$ mpc; cf., footnote~\ref{footnote-hst1}), the predicted IC 
spectrum has a complex temporal and spectral behavior. In the absence of 
detailed contemporaneous measurements, its possible role in the 
production of the LAT observed MeV/GeV emission is unclear.

Continued LAT monitoring of M87 coordinated with multi-wavelength 
observations can extend the current study of `quiescent' emission to 
possible flaring, in order to further address the physics of the 
radiation zone. While the extragalactic $\gamma$-ray sky is dominated by 
blazars \citep{har99,lbas}, this optimistically indicates an emerging 
population of $\gamma$-ray radio galaxies.  Other examples, including 
the few possible associations with EGRET detections like, NGC~6251 
\citep{muk02} and 3C~111 \citep{sgu05,har08} await confirmation with the 
LAT, and more radio galaxies are expected to be detected at lower 
fluxes. This holds great promise for systematic studies of relativistic 
jets with a range of viewing geometries in the high energy $\gamma$-ray 
window opened up by the $Fermi$-LAT.

\acknowledgments

The $Fermi$ LAT Collaboration acknowledges generous ongoing support from 
a number of agencies and institutes that have supported both the 
development and the operation of the LAT as well as scientific data 
analysis.  These include the National Aeronautics and Space 
Administration and the Department of Energy in the United States, the 
Commissariat \`a l'Energie Atomique and the Centre National de la 
Recherche Scientifique / Institut National de Physique Nucl\'eaire et de 
Physique des Particules in France, the Agenzia Spaziale Italiana and the 
Istituto Nazionale di Fisica Nucleare in Italy, the Ministry of 
Education, Culture, Sports, Science and Technology (MEXT), High Energy 
Accelerator Research Organization (KEK) and Japan Aerospace Exploration 
Agency (JAXA) in Japan, and the K.~A.~Wallenberg Foundation, the Swedish 
Research Council and the Swedish National Space Board in Sweden.

Additional support for science analysis during the operations phase is 
gratefully acknowledged from the Istituto Nazionale di Astrofisica in 
Italy.

C.C.C. was supported by an appointment to the NASA Postdoctoral Program 
at Goddard Space Flight Center, administered by Oak Ridge Associated 
Universities through a contract with NASA. Support from NASA grants 
GO8-9116X and GO9-0108X (D.E.H., F.M.) and the Foundation BLANCEFLOR 
Boncompagni-Ludovisi, n'ee Bildt (F.M.) are acknowledged. This research 
has made use of data from the MOJAVE database that is maintained by the 
MOJAVE team \citep{lis09}. We thank F.~Owen for providing the VLA 90cm 
image.

{}

\end{document}